\documentclass[aps,pra,twocolumn,showpacs,superscriptaddress]{revtex4}
\usepackage{graphicx}
\usepackage{amsmath}
\usepackage{amssymb}

\input{epsf}
\begin{document}

\title{Universal four-body states in heavy-light mixtures with
positive scattering length}

\author{D. Blume}
\affiliation{Department of Physics and Astronomy,
Washington State University,
  Pullman, Washington 99164-2814, USA}
\affiliation{ITAMP, Harvard-Smithsonian Center for Astrophysics,
60 Garden Street, Cambridge, Massachusetts 02138, USA}

\date{\today}

\begin{abstract}
The number of four-body states known to behave universally
is small. This work adds a new class of four-body
states to this relatively short list.
We predict the existence of a universal
four-body bound state for heavy-light mixtures consisting of three
identical heavy fermions and a fourth distinguishable lighter
particle with mass ratio $\kappa \gtrsim 9.5$
and short-range interspecies interaction
characterized by a positive $s$-wave scattering length.
The structural properties of these universal states
are discussed and finite-range effects are analyzed.
The bound states can be experimentally realized 
and probed utilizing ultracold
atom mixtures.
\end{abstract}


\maketitle

Universality plays an important role in nearly all
areas of physics and allows one to connect phenomena governed by vastly
different energy and length scales.
A simple class of universal states
consists of two-body bound states whose size is much larger than any
other length scale in the problem.
Prominent examples include
diatomic Feshbach molecules~\cite{chin}, which are
nowadays created routinely in cold atom laboratories
around the world, and di-mesons
such as the charmonium
resonance near 3870~MeV~\cite{braaten}.
The former have a binding energy of order $10^{-10}$~eV while 
the latter have a binding 
energy of order $0.5 \times 10^6$~eV.
Yet, once expressed in terms of the two-body $s$-wave scattering length
$a_s$, the binding energy can be written, to a very good approximation,
as $E_2^{\rm{ZR}} \approx -\hbar^2/(2 \mu a_s^2)$ in both cases;
here, $\mu$ is the reduced mass of the constituents
(the two atoms and two mesons, respectively).

Although the concept of universality has been extended successfully to
three- and higher-body 
systems~\cite{chrisPhysToday,rudiTrend,braatenReview,platter1,hanna,platter2,vonstecher,vonstecher2,deltuva,grimmnature,grimm,hulet,zaccanti,fivebody,petrov,esry1,esry2,kart07,cast11,levi12}, the list of examples,
particularly for few-body systems consisting of more than
$n=3$ constituents, is still comparatively
small. Most notably, three- and four-body physics has been
investigated in the context of Efimov physics.
The three-body Efimov effect~\cite{braatenReview}, i.e.,
the existence of infinitely many geometrically spaced
three-body bound states, can occur when the $s$-wave scattering length
is much larger than the range of the two-body potential. This at first
sight purely academic scenario can be realized in cold atom
experiments by tuning the $s$-wave scattering length in the
vicinity of a Fano-Feshbach resonance through application
of an external magnetic or optical field~\cite{chin,grimmnature}.
In the four-body sector, Efimov physics can occur via two different
routes, as a true four-body Efimov effect~\cite{cast11}
or as four-body states universally tied to three-body Efimov
states~\cite{platter1,vonstecher,platter2}.
In either case, the description 
of the Efimov scenario requires two parameters,
the $s$-wave scattering length and a three-body 
parameter~\cite{braatenReview}.
The latter may be determined by the characteristic length of
the underlying two-body potential~\cite{chrisPRL}
(a scenario that has been suggested to apply to cold atom systems)
or may depend on true three-body physics 
(a scenario more likely to apply to nuclear systems).

This Letter reports on a new class of universal four-body states,
predicted to exist---just as Efimov states---in three spatial dimensions
that are fully determined by the two-body $s$-wave scattering length
$a_s$.
As such, they are fundamentally different from Efimov states, 
which depend on two parameters. Another crucial distinction is
that the states considered here have continuous scale invariance 
while Efimov states exhibit discrete scale invariance.
The universal four-body bound states exist in heavy-light mixtures
that consist of three identical 
heavy fermions and a fourth distinguishable particle,
which interacts with the heavy 
particles through a short-range two-body potential with positive
$s$-wave scattering length $a_s$.
We find that the four-body bound states 
exist for mass ratios $\kappa$ larger than $\kappa_{c,4} \approx 9.5$.
For effectively two- or one-dimensional confinement,
the universal tetramers are expected to be more strongly
bound than in three spatial dimensions.
In fact, universal tetramers under quasi-two-dimensional confinement
have very recently been predicted to exist for 
$\kappa \gtrsim 5$~\cite{levi12}.
Just as the three-body bound states for positive
$a_s$ are connected to Efimov states (which exist, in the zero-range limit,
for 
$\kappa > 13.607$)~\cite{kart07,petrov,esry1,esry2,uedagroup,unpublished}, 
the universal four-body states 
predicted here are expected to be connected to four-body
Efimov states, 
which have been predicted to exist for 
$13.384 < \kappa < 13.607$~\cite{cast11}.
We analyze the dependence of 
the binding energy on the range of the underlying two-body
interaction potential and interpret our findings
employing hyperspherical coordinates.
The universal four-body bound states discussed here are not only
interesting from the 
few-body point of view but also have important implications
for the many-body phase diagram of 
heavy-light mixtures~\cite{iskin1,iskin2,nishida,mathy}
that can be realized with cold 
atoms~\cite{coldatoms1,coldatoms2,coldatoms3,coldatoms4}, 
electrons~\cite{electrons} 
and quarks.

Our starting point is the non-relativistic Hamiltonian $H$
in free space
for $n-1$ identical 
heavy fermions of mass $M$ and a single 
distinguishable light particle
of mass $m$,
\begin{eqnarray}
\label{eq_ham}
H = \sum_{j=1}^{n-1} \frac{-\hbar^2}{2M} \nabla_{\vec{r}_j}^2  
+\frac{-\hbar^2}{2m} \nabla_{\vec{r}_n}^2  + 
\sum_{j=1}^{n-1} V_{\rm{tb}}(r_{jn}),
\end{eqnarray}
where
\begin{eqnarray}
\label{eq_vint}
V_{\rm{tb}}(r_{jn}) = 
-V_0 \exp \left[- r_{jn}^2/(2 r_0^2) \right].
\end{eqnarray}
Here, $\vec{r}_j$ denotes the position vector of the $j$th
particle and $r_{jk}$ the interparticle distance,
$r_{jk}=| \vec{r}_j - \vec{r}_k|$.
The interaction between the heavy and light particles
is described by the Gaussian potential $V_{\rm{tb}}$ with
depth $V_0$ and range $r_0$.
In the following, we are interested in the regime where
the two-body free-space 
$s$-wave scattering length $a_s$ of $V_{\rm{tb}}$ is positive and 
$r_0 \ll a_s$.
Throughout, we express lengths in units of $a_s$ and energies
in units of $|E_2^{\rm{ZR}}|$, where $E_2^{\rm{ZR}}$ denotes the relative
$s$-wave energy of the two-body system with zero-range interactions
(realized when $r_0 \rightarrow 0$),
$E_2^{\rm{ZR}} = -\hbar^2/(2\mu a_s^2)$ with
$\mu=Mm/(M+m)$.
For a given $r_0/a_s$, we adjust the depth $V_0$ such that the
two-body potential supports a single $s$-wave bound state.

To determine the eigenstates and eigenenergies of $H$,
we separate off the center-of-mass degrees of freedom and expand the
relative wave function in terms of explicitly correlated 
Gaussians~\cite{cgbook}.
To construct basis functions
with good total relative angular 
momentum $L$, projection quantum number $M_L$,
and parity $\Pi$, 
we employ the global vector approach~\cite{suzu08,debraj},
which has been used extensively to describe nuclear systems.
Throughout, we limit ourselves to states with $M_L=0$.
The anti-symmetry of the basis functions with respect to the
exchange of pairs of identical
fermions is ensured through the application of an
anti-symmetrizer.
The parameters of the explicitly correlated Gaussian basis functions are 
optimized semi-stochastically. According to the generalized Ritz
variational principle~\cite{cgbook}, the approach yields variational upper
bounds for the eigenenergies of the ground and excited states.
Our stochastic variational
calculations for $n=3$ and $4$ reported below employ
up to 1000 and 3500 
basis functions, respectively.

\begin{figure}
\vspace*{+.5cm}
\includegraphics[angle=0,width=70mm]{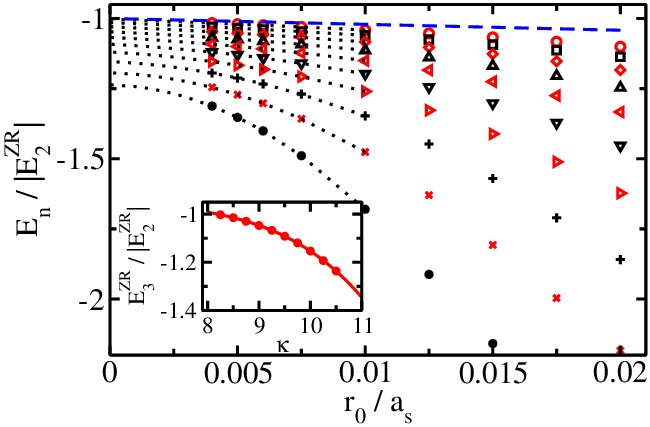}
\vspace*{0.cm}
\caption{(Color online)
Scaled energies as a function of $r_0/a_s$.
The dashed line shows $E_2/|E_2^{\rm{ZR}}|$.
The symbols show $E_3/|E_2^{\rm{ZR}}|$ 
for $\kappa=8.25$ to $\kappa=10.5$ (top to bottom),
in steps of $0.25$;
$E_3$ is determined by the stochastic
variational approach.
Dotted lines show three-parameter fits.
Inset: 
Symbols show the extrapolated three-body zero-range energy $E_3^{\rm{ZR}}$
as a function of $\kappa$. The solid line shows a four-parameter fit.
}\label{fig1}
\end{figure}

We first consider the $n=3$
system with $L^{\Pi}=1^-$ symmetry.
Employing zero-range
$s$-wave interactions, a universal trimer state has been predicted
to exist for $\kappa_{c,3} \gtrsim 8.173$~\cite{kart07}.
A second universal trimer state has been predicted to be supported
for $\kappa_{c,3}^* \gtrsim 12.917$~\cite{kart07}.
Symbols in Fig.~\ref{fig1} show the relative energy 
$E_3$ of the energetically lowest-lying three-body state
with $1^-$ symmetry,
calculated by the stochastic variational approach and
normalized by $|E_2^{\rm{ZR}}|$, as a function
of $r_0/a_s$ for various mass ratios ($\kappa = 8.25-10.5$,
where $\kappa=M/m$).
As expected, the trimer energy becomes
more negative with increasing $\kappa$ for a 
fixed $r_0/a_s$. Moreover, the trimer energies approach the zero-range 
limit from below, with the range dependence becoming larger
with increasing $\kappa$.
For comparison, the dashed line shows the quantity
$E_2/|E_2^{\rm{ZR}}|$ as a function of $r_0/a_s$;
here, $E_2$ denotes the relative two-body energy.
Due to the scaling chosen, the dimer energy is independent of the 
mass ratio.
The dependence of the dimer energy on $r_0$ is smaller
than that of the trimer energy.
Dotted lines in Fig.~\ref{fig1} show three-parameter fits to the
three-body
energies with $r_0/a_s \le 0.01$~\cite{notefit}.
The symbols in the inset of Fig.~\ref{fig1} show the extrapolated zero-range
energies 
$E_3^{\rm{ZR}}$ 
of the trimer, scaled by $|E_2^{\rm{ZR}}|$, as a function
of the mass ratio $\kappa$. 
The solid line shows a fit of the
quantity $E_3^{\rm{ZR}}/|E_2^{\rm{ZR}}|$ to a fourth-order polynomial.
Our fit predicts that the trimer becomes unbound with respect to the dimer
for $\kappa_{c,3} \approx 8.20$, which compares
favorably with the $\kappa_{c,3}$ value
of $8.173$ 
determined for zero-range interactions~\cite{kart07}.
This good agreement demonstrates that the stochastic variational
approach employed in this
work  is capable of  accurately
describing universal few-body bound states.
It should be noted that non-universal trimer states can, 
at least in principle, exist 
for $\kappa \gtrsim 8.6$~\cite{petrov,nishida,wernerandcastin}.
Whether or not non-universal states exist depends on the
details of the underlying two-body potential. For the Gaussian model
potential considered here, it was shown earlier~\cite{blume1,blume2} 
that non-universal three-body
physics comes into play for mass ratios larger
than those considered here.

We now discuss the energetics of the four-body system.
\begin{figure}
\vspace*{+.5cm}
\includegraphics[angle=0,width=70mm]{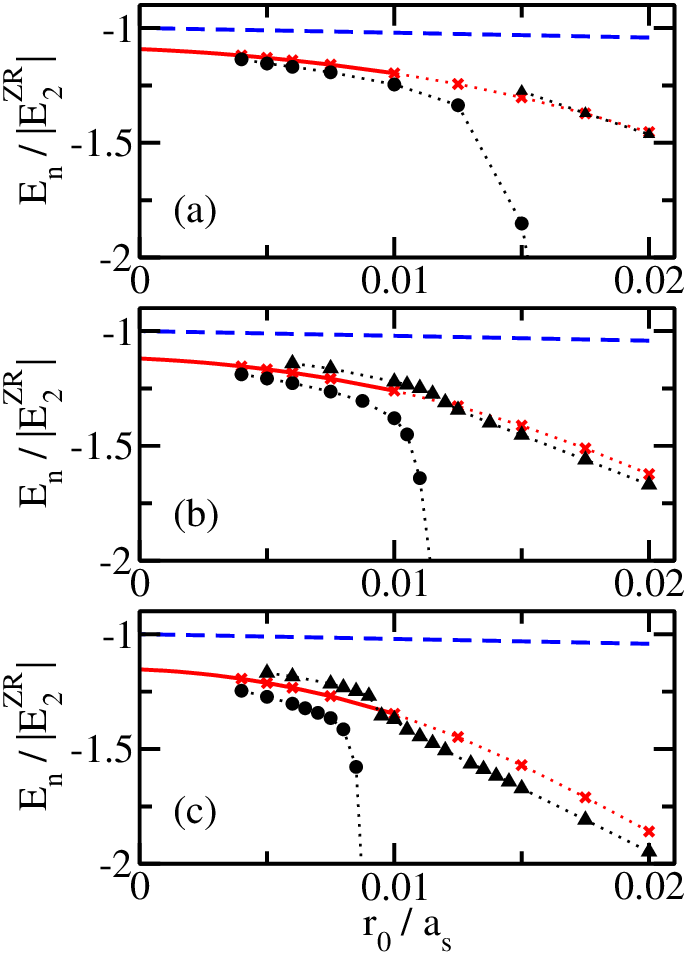}
\vspace*{0.cm}
\caption{(Color online)
Scaled energies as a function of $r_0/a_s$ for (a) $\kappa=9.5$,
(b) $\kappa=9.75$, and
(c) $\kappa=10$.
The dashed lines show $E_2/|E_2^{\rm{ZR}}|$ 
while the crosses (stochastic variational energies) 
and solid lines (fit) show
$E_3/|E_2^{\rm{ZR}}|$ 
(these energies are also shown in Fig.~\protect\ref{fig1}).
The circles and triangles show $E_4/|E_2^{\rm{ZR}}|$ for the energetically
lowest-lying and second lowest-lying four-body states,
respectively.
Dotted lines serve as a guide to the eye.
}\label{fig2}
\end{figure}
Circles 
and triangles 
in Fig.~\ref{fig2} 
show the quantity $E_4/|E_2^{\rm{ZR}}|$ for, respectively, 
the energetically lowest-lying 
and second lowest-lying states
of the four-body system with $L^{\Pi}=1^+$
symmetry
for (a) $\kappa=9.5$, (b) $\kappa=9.75$ and (c)
$\kappa=10$ as a function of $r_0/a_s$.
The four-body energies are
obtained by the stochastic variational
approach. Dotted lines are shown as a guide to the eye.
For comparison, the dashed lines in Fig.~\ref{fig2} show the quantity
$E_2/|E_2^{\rm{ZR}}|$, and the crosses and solid lines show
the quantity $E_3/|E_2^{\rm{ZR}}|$ (the symbols show the 
stochastic variational energies and the solid line
shows the fit; see also Fig.~\ref{fig1}). 
The ground state energy of the four-body system lies below that 
of the three-body system for 
small $r_0/a_s$. For $\kappa=9.5$, $9.75$ and $10$,
the ground state energy of the four-body system ``dives down'' around 
$r_0/a_s \approx 0.015$, $0.011$ and $0.008$, respectively. 
In this regime, the
four-body state acquires
non-universal characteristics. 
For slightly larger $r_0/a_s$, the energy of the first excited state
drops below the energy of the trimer and then ``traces''
the three-body energy~\cite{footnoteenergy}.
We refer to the
feature 
where the four-body system acquires a new bound state as resonance-like
feature.
The existence and characteristics of the resonance-like feature
depend on the details of the two-body interaction model employed.
Away from the resonance-like feature, 
the four-body energy shows a very similar range
dependence as the three-body energy, suggesting that the four-body energy is
roughly a constant multiple of the three-body energy.
Moreover, it is clear from Fig.~\ref{fig2} that the ratio between
the four- and three-body energy increases with increasing mass 
ratio~\cite{footnotelargerkappa}.
A precise extrapolation of the four-body energies to the zero-range 
limit is challenging  for two reasons: {\em{(i)}}
Numerical issues limit our calculations to $r_0/a_s \gtrsim 0.004$.
{\em{(ii)}} The existence of the resonance-like feature prevents us to 
perform unambiguous fits.
We estimate that 
the four-body system becomes 
bound around $\kappa_{c,4}=9.5$.

To provide further
evidence that the four-body states are---away from the resonance-like
feature---universal,
we
analyze the hyperradial density $P(R)$.
The hyperradius $R$,
\begin{eqnarray}
\label{eq_rhyper}
R= \mu^{-1/2}
\sqrt{
\sum_{j=1}^{n-1} M (\vec{r}_j-\vec{R}_{\rm{cm}})^2+ 
m (\vec{r}_n-\vec{R}_{\rm{cm}})^2
}, 
\end{eqnarray}
provides  a measure of the system 
size~\cite{ritt11}.
A small hyperradius implies that all $n$ particles are close
together while a large hyperradius implies that two or more
particles are far away from each other.
In Eq.~(\ref{eq_rhyper}), $\vec{R}_{\rm{cm}}$ 
denotes the center-of-mass vector of the $n$-body system.
The hyperradial density $P(R)$,
normalized such that $\int_0^{\infty} P(R) dR= 1$, indicates
the likelihood of finding the $n$-particle system with a given $R$.
We calculate the hyperradial densities
as well as other structural properties
by sampling the $n$-particle
density obtained by the stochastic variational approach
via a Metropolis walk~\cite{blume3}.

\begin{figure}
\vspace*{+.5cm}
\includegraphics[angle=0,width=70mm]{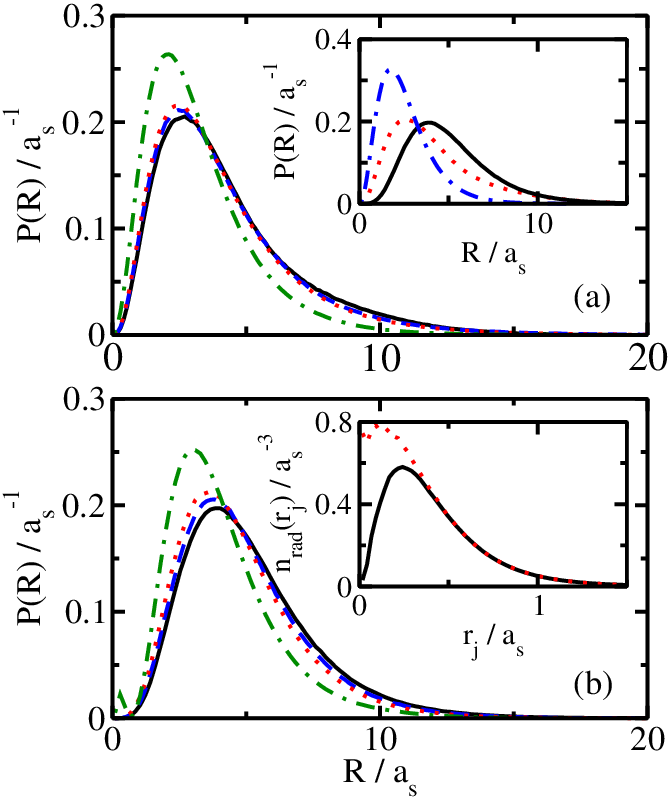}
\vspace*{0.cm}
\caption{(Color online)
Structural properties.
(a) Solid, dashed, dotted and dash-dotted lines show the hyperradial density
$P(R)$
of the energetically lowest-lying state
for $n=3$,
$\kappa=8.5$ and $r_0/a_s=0.004$, $0.005$, $0.006$
and $0.015$, respectively.
Inset: Dotted, dash-dotted and solid lines show the hyperradial density for
$n=3$ ($\kappa=8.5$ and $r_0/a_s=0.004$),
$n=3$ ($\kappa=9.75$ and $r_0/a_s=0.004$),
and
$n=4$ ($\kappa=9.75$ and $r_0/a_s=0.004$), respectively.
(b) Solid, dashed, dotted and dash-dotted lines show $P(R)$
for the four-body system with 
$\kappa=9.75$ and $r_0/a_s=0.004$, $0.005$, $0.006$ and
$0.015$, respectively.
For $r_0/a_s=0.004, 0.005$ and $0.006$, 
the energetically lowest-lying state is considered.
For $a_s/r_0=0.015$,
the energetically second-lowest lying state is considered.
Inset: 
Solid and dotted lines show the radial density 
$n_{\rm{rad}}(r_j)$ for the heavy and 
light particles, respectively, of the $n=4$ system
with $\kappa=9.75$, $r_0/a_s=0.004$ and $1^+$ symmetry.
}\label{fig3}
\end{figure}

Figures~\ref{fig3}(a) and \ref{fig3}(b) show the hyperradial densities
$P(R)$ for $n=3$ ($\kappa=8.5$ and $L^{\Pi}=1^-$)
and $n=4$ ($\kappa=9.75$ and $L^{\Pi}=1^+$) for various 
$r_0/a_s$~\cite{footnotehyperradialdensity}.
For these mass ratios, the three- and four-body systems support 
very weakly-bound states.
To allow for a direct comparison, dotted and solid 
lines in the inset of Fig.~\ref{fig3}(a) show the hyperradial densities for
$n=3$ ($\kappa=8.5$) and $n=4$
($\kappa=9.75$) for $r_0/a_s=0.004$.
The hyperradial densities for $n=3$ with $\kappa=8.5$
and $n=4$ with $\kappa=9.75$ agree qualitatively.
They
have a small amplitude for $R/a_s \ll 1$, 
peak around $R/a_s=2$ and fall off exponentially
for $R \gg a_s$ for all $r_0/a_s$ considered. 
For fixed $\kappa$, the hyperradial densities 
move smoothly ``outward''
with decreasing $r_0/a_s$.
Importantly, the hyperradial density has vanishingly small amplitude 
not only when $R \approx r_0$ but also
for notably larger $R$ values~\cite{footnote_smallR}. 
For the three-body system,
this is consistent with the
hyperradial density obtained within
the zero-range framework~\cite{kart07,footnotehyperradial}, confirming
that the three-body states considered are fully
universal, i.e., fully determined by $a_s$.
The qualitatively similar behavior of the $n=3$ and 4 hyperradial densities 
for similarly weakly-bound states
provides, combined with the energetics, strong evidence
that the four-body states are also universal.

The dash-dotted line
in the inset of Fig.~\ref{fig3}(a) shows the hyperradial density of
the three-body system
for $\kappa=9.75$ and 
$r_0/a_s=0.004$.
The three-body system is more 
tightly bound than the four-body system
with the same $\kappa$ and $r_0/a_s$ (solid line).
In a naive picture, one may imagine that the four-body system is 
comprised of a trimer with a fourth atom 
loosely attached to the trimer.
Structures like this have been observed for the excited 
tetramer 
state
attached to the Efimov trimer
state comprised of three identical bosons~\cite{vonstecher,vonstecher2}.
Our analysis of the pair distribution functions and 
radial densities indicates that the situation for 
the tetramers considered here
is different.
The structural properties of the tetramer of a given $\kappa$
loosely resemble those
of the trimer with smaller $\kappa$ but
comparable binding energy.
Solid and dotted lines in
the inset of Fig.~\ref{fig3}(b) show the radial density
$n_{\rm{rad}}(r_j)$, normalized such that
$4 \pi \int_0^{\infty} n_{\rm{rad}}(r_j) r_j^2 dr_j=1$,
for the heavy and light particles of the $n=4$ system;
the position vector $\vec{r}_j$, $j=1,\cdots,n$,
is measured with respect to $\vec{R}_{\rm{cm}}$ and
$r_j=|\vec{r}_j|$.
For large $r_j$, the radial densities of the heavy particles
and the light particle are nearly indistinguishable. For small $r_j$, 
$n_{\rm{rad}}$ goes to zero for the heavy particles but has an
appreciable amplitude for the light particle, suggesting
that the light particle is ``shared'' among the heavy particles.

In summary,
we analyzed heavy-light mixtures
in three spatial dimensions, where the heavy-light pairs
interact through
short-range potentials with positive $s$-wave scattering lengths.
Despite the Pauli exclusion principle, which acts as an effective
repulsion between the identical heavy fermions, 
the four-body system consisting of three heavy particles and one light particle
supports a universal bound state if 
the mass ratio between the heavy and 
light particles is larger than about $9.5$.
The light particle acts as a mediator that ``glues''
the four-body system together, just as
electrons in H$_2^+$ or H$_2$
glue together the protons by way of the exchange
interaction~\cite{bransden}.
Although the three-body energy shows a fairly
strong dependence on $r_0$, 
we found that the ratio $E_4/E_3$ is, away from the resonance-like feature,
roughly constant for a fairly wide range of $r_0/a_s$
values,
suggesting that the universal four-body states can be observed in 
cold atom experiments with current technologies.
The existence of universal tetramer states opens the possibility
to search for novel tetramer phases in many-body systems,
promising a rich phase diagram of heavy-light mixtures
on the positive scattering length side.
In the future, it will be interesting to investigate how 
the universal four-body states discussed here are affected by
non-universal three- and four-body states and how these states are
connected  to Efimov tetramers that have been predicted to
exist for $13.384 < \kappa < 13.607$~\cite{cast11}.

{\bf{Acknowledgments:}}
DB thanks Javier von Stecher for suggesting to look at
heavy-light mixtures with
positive scattering length and Seth Rittenhouse for
fruitful discussions.
Support by the NSF through
grants PHY-0855332 and PHY-1205443
is gratefully acknowledged.
This work was additionally supported by the National Science
Foundation through a grant for the Institute for
Theoretical Atomic, Molecular and Optical Physics
at Harvard University and Smithonian Astrophysical Observatory.


\begin{thebibliography}{100}
\bibitem{notefit}
We fit a polynomial
containing a constant, a linear and
a quadratic term to the
quantity $E_3/|E_2|$, where $E_2$ denotes the range-dependent 
two-body energy.
\bibitem{footnoteenergy}
For $\kappa=9.5$,
the energy of the four-body state with $1^+$ symmetry lies
slightly below the energy of the three-body state with $1^-$ symmetry
for $r_0/a_s=0.02$ but not for $r_0/a_s=0.0175$ and $0.015$.
\bibitem{footnotelargerkappa}
We also performed calculations for $\kappa=10.25$ and $10.5$.
For these mass ratios, the finite-range 
effects of the three-body system are already
fairly large (see Fig.~\protect\ref{fig1}).
For the $r_0/a_s$ considered,
we also 
observe that the four-body energy is, away from the resonance-like feature
(which moves to smaller $r_0/a_s$ with increasing $\kappa$),
roughly a constant multiple of the three-body energy.
\bibitem{footnotehyperradialdensity}
For $n=4$ and $r_0/a_s=0.015$,
the hyperradial density of the 
energetically second lowest-lying state with $1^+$ symmetry
is shown [this is the state whose energy is
shown by a triangle
in Fig.~\ref{fig2}(b)]; for all other cases,
the hyperradial density of the energetically lowest-lying state is shown.
The 
wave function 
of the 
energetically second lowest-lying state 
has a node in the hyperradial coordinate, which leads to the
vanishing
of the hyperradial density around
$r_0/a_s \approx 0.5$.
\bibitem{footnote_smallR}
For $n=3$ and $R=10r_0$ ($r_0/a_s=0.004$ and $\kappa=8.5$),
the amplitude of the hyperradial density is more
than 1000 times smaller than the corresponding hyperradial peak density.
For $n=4$ and $R=50r_0$ ($r_0/a_s=0.004$ and $\kappa=9.75$),
the amplitude of the hyperradial density is more
than 1000 times smaller than the corresponding hyperradial
peak density.
\bibitem{footnotehyperradial}
Reference~\protect\cite{kart07} does not
present hyperradial densities for the three-body system. 
However, the hyperradial density for the three-body system
within the zero-range framework
is readily obtained using the results of Ref.~\protect\cite{kart07}.
\end{thebibliography}
\end{document}